\documentclass[sort&compress,5p,floatfix]{elsarticle}

\usepackage{amssymb}
\usepackage{bm}
\usepackage{amsmath}
\usepackage{color}
\definecolor{darkblue}{RGB}{0,0,196}
\definecolor{darkred}{RGB}{196,0,0}
\usepackage[colorlinks=true,linktocpage=true,linkcolor=darkblue,citecolor=darkblue,urlcolor=darkblue]{hyperref}
\usepackage{widetext}

\def\be{\begin{equation}}
\def\ee{\end{equation}}
\def\ba{\begin{eqnarray}}
\def\ea{\end{eqnarray}}

\journal{Physics Letters B}

\begin{document}

\begin{frontmatter}



\title{Bottomonium suppression and elliptic flow from real-time quantum evolution}


\author{Ajaharul Islam and Michael Strickland}
\ead{aislam2@kent.edu and mstrick6@kent.edu}

\address{Department of Physics, Kent State University, Kent, OH 44242, United States}

\begin{abstract}
We compute the suppression and elliptic flow of bottomonium using real-time solutions to the Schr\"{o}dinger equation with a realistic in-medium complex-valued potential.  To model the initial production, we assume that, in the limit of heavy quark masses, the wave-function can be described by a lattice-smeared (Gaussian) Dirac delta wave-function.  The resulting final-state quantum-mechanical overlaps provide the survival probability of all bottomonium eigenstates.  Our results are in good agreement with available data for $R_{AA}$ as a function of $N_{\rm part}$ and $p_T$ collected at \mbox{$\sqrt{s_{\rm NN}} =$ 5.02 TeV}.  In the case of $v_2$ for the various states, we find that the path-length dependence of $\Upsilon(1s)$ suppression results in quite small $v_2$ for $\Upsilon(1s)$.  Our prediction for the integrated elliptic flow for $\Upsilon(1s)$ in the $10{-}90$\% centrality class is $v_2[\Upsilon(1s)] = 0.0026 \pm 0.0007$.  We additionally find that, due to their increased suppression, excited bottomonium states have a larger elliptic flow and we make predictions for $v_2[\Upsilon(2s)]$ and $v_2[\Upsilon(3s)]$ as a function of centrality and transverse momentum.  Similar to prior studies, we find that it is possible for bottomonium states to have negative $v_2$ at low transverse momentum.
\end{abstract}

\cortext[cor1]{Corresponding authors}






\end{frontmatter}

\section{Introduction}

Relativistic heavy-ion collision experiments at Brook-haven National Laboratory's (BNL) Relativistic Heavy Ion Collider (RHIC) and the European Organization for Nuclear Research's (CERN) Large Hadron Collider (LHC) study the behavior of matter subject to extreme conditions.  The goal of these experiments is to create and study the properties of the Quark-Gluon Plasma (QGP) which is expected to be produced when the energy density of matter exceeds approximately \mbox{1 GeV/fm$^3$}.  Detailed lattice studies have demonstrated that quantum chromodynamics (QCD) has a pseudo-critical temperature of approximately $T_{\rm pc} \simeq 155$ MeV \cite{Bazavov:2013txa,Borsanyi:2016bzg}.  Since the QGP is a color-ionized phase of matter, it is expected to strongly affect the propagation of both individual partons and hadronic bound states.

Hadrons composed of light quarks are expected to disassociate at temperatures around, or just above, $T_{\rm pc}$.    For heavy-quarkonium bound states, such as the $J/\psi$ and $\Upsilon$, however, it was predicted in the late 1980s that such states could survive into the QGP phase due to their large binding energy~\cite{Matsui:1986dk,Karsch:1987pv,PhysRevD.70.054507,PhysRevC.70.021901}.  QCD-based model and lattice gauge theory calculations have found that the $J/\psi$ and $\Upsilon$ disassociation temperatures are approximately 250-400 MeV and 450-700 MeV, respectively~\cite{Andronic:2015wma,Mocsy:2013syh,Brambilla:2010cs,Brambilla:2010vq}.  Before one reaches these high temperatures, however, one expects there to be partial suppression of heavy-quarkonium bound states due to in-medium breakup processes related to, for example, Landau damping and singlet-octet transitions.  Because of this, one can use heavy quarkonium bound states as an internally generated probe of the QGP, with their survival probabilities depending on QGP properties such as its initial temperature and the size of expected non-equilibrium deviations.

In the pioneering papers of Karsch, Matsui, and Satz (KMS) they made the first predictions that heavy quarkonia would ``melt'' in the QGP~\cite{Matsui:1986dk,Karsch:1987pv}.  These studies were based on a non-relativistic potential-based model and the disassociation temperature of states were obtained by finding when the binding energy of the state goes to zero or $\langle r \rangle \rightarrow \infty$.  Such a non-relativistic treatment is justified by the fact that, as the mass of the heavy-quark increases, its velocity inside the bound state decreases and, for sufficiently heavy quarks, e.g. bottom quarks, one can construct a non-relativistic effective field theory and then solve the non-relativistic Schr\"odinger equation with the resulting in-medium potential.  This process can be made more formal using effective field theory methods to integrate out different energy/momentum scales, resulting in potential-based non-relativistic QCD (pNRQCD) \cite{PhysRevD.21.203,Lucha:1991vn,Pineda:1997bj,Brambilla:1999xf,Brambilla:2010xn}.  

Based on high-temperature quantum field theory calculations and complementary effective field theory calculations, it is now known that the in-medium heavy-quark potential is complex-valued, with the imaginary part being related to the in-medium breakup rate of heavy-quark bound states \cite{Laine:2006ns,Dumitru:2007hy,Brambilla:2008cx,Burnier:2009yu,Dumitru:2009fy,Dumitru:2009ni,Margotta:2011ta,Du:2016wdx,Nopoush:2017zbu,Guo:2018vwy}.   This imaginary part has been shown to be related to gluon dissociation or parton free dissociation of the states in Refs.~\cite{Brambilla:2011sg,Brambilla:2013dpa}.  The imaginary part of the potential makes the quantum evolution non-unitary (non-Hermitian Hamiltonian), which can be understood in the context of open quantum systems in which there is a heavy-quark bound state coupled to a thermal heat bath \cite{Akamatsu:2011se,Akamatsu:2012vt,Akamatsu:2014qsa,Katz:2015qja,Brambilla:2016wgg,Kajimoto:2017rel,Brambilla:2017zei,Blaizot:2017ypk,Blaizot:2018oev,Yao:2018nmy}.  We note that, in the context of transport models, in-medium breakup is also included in studies of bottomonium and charmonium suppression  \cite{Grandchamp:2005yw,Rapp:2008tf,Emerick:2011xu,Du:2017qkv,Yao:2018sgn,Du:2019tjf,Yao:2020xzw}.

In this paper, we focus on bottomonium states and present a model called Heavy Quarkonium Quantum Dynamics (HQQD) in which we solve the time-dependent Schr\"{o}dinger equation with a complex in-medium potential for a large set of Monte-Carlo-sampled bottomonium wave-packet trajectories.  We consider only \mbox{$\sqrt{s_{NN}} = 5.02$ TeV} Pb-Pb collisions herein.  For each trajectory, the states are in a quantum linear superposition and we extract the survival probability of a given state by computing the quantum-mechanical overlap of the state's vacuum eigenstate with the in-medium evolved quantum wave-function. We do not include explicit time-dependent noise contributions in the potential and, as a result, for singlet evolution the obtained survival probabilities correspond to those associated with the average wave-function (averaged over thermal fluctuations) \cite{Kajimoto:2017rel}.  Although this is an approximation, it is a very reasonable starting point for updated phenomenological studies.  Many phenomenological studies presented in the past have used this approximation to solve for the evolution of the average wave-function, however, they additionally made use of the adiabatic approximation which allows one to compute the instantaneous breakup rate for a given state from time-independent solutions to the Schr\"{o}dinger equation \cite{Margotta:2011ta,Strickland:2011mw,Strickland:2011aa,Strickland:2012cq,Krouppa:2015yoa,Krouppa:2016jcl,Krouppa:2017jlg,Jaiswal:2017dxp,Bhaduri:2018iwr,Bhaduri:2020lur}.  This approximation throws out potentially important physics such as quantum state mixing due to the time-dependent in-medium potential.  

In a previous paper \cite{Boyd:2019arx}, we made a preliminary investigation of the effects of relaxing the adiabatic approximation, finding that there were potentially important effects on the survival probability of the states.  Herein, we turn this approach into a more complete phenomenological framework, which can be used for comparisons with experimental data.  To do this, we make use of the output of a 3+1D anisotropic hydrodynamics code which has been tuned to reproduce a large set of soft hadronic observables in $\sqrt{s_{NN}} = 5.02$ TeV collisions \cite{Florkowski:2010cf,Martinez:2010sc,Bazow:2013ifa,Alqahtani:2017jwl,Alqahtani:2017tnq,Alqahtani:2017mhy,Almaalol:2018gjh,mubarakforth}.  After computing each state's survival probability, we then take into account late-time feed down of excited states using vacuum branching ratios available from the Particle Data Group \cite{pdg}.  We find that our HQQD results are in quite reasonable agreement with available data given current uncertainties, however, some quantitative differences remain which motivate going beyond the methods used herein to more fully include the effects of in-medium thermal noise, initial production in octet states, and singlet-octet transitions.

\section{Methodology}

We solve the real-time Schr\"{o}dinger equation with a complex-valued potential of the form $V(r) = V_R(r) + iV_I(r)$.  We assume that, in vacuum, the singlet heavy-quarkonium potential is given by a Cornell potential with finite string-breaking distance
\be
V_{\rm vac}(r) =
\begin{cases}  
-\frac{a}{r} + \sigma r &\mbox{if } r \leq r_{\rm SB} \\
-\frac{a}{r_{\rm SB} } + \sigma r_{\rm SB}   & \mbox{if } r > r_{\rm SB}
\end{cases} \, ,
\label{eq:vvac}
\ee
where \mbox{$a = 0.409$} is the effective coupling, \mbox{$\sigma = 0.21$~GeV$^2$} is the string tension, and \mbox{$r_{\rm SB}  =$ 1.25 fm} is the string breaking distance.  With this tuning of the vacuum potential, and assuming $M_b = 4.7$ GeV, we obtain vacuum masses of \mbox{$\{9.46,10.0,9.88,10.36,10.25,10.13\}$ GeV} for $\Upsilon(1s)$, $\Upsilon(2s)$, $\chi_b(1p)$, $\Upsilon(3s)$, and $\chi_b(2p)$, respectively.

The real-part of the finite-temperature singlet quark-antiquark potential is taken to be given by the internal-energy associated with the Karsch-Mehr-Satz (KMS) potential  \cite{Strickland:2011mw,Strickland:2011aa}
\ba
V_{\rm KMS}(r) &=& -\frac{a}{r}(1 + m_D r)e^{-m_D r} \nonumber \\
&& + \frac{2\sigma}{m_D}[1-e^{-m_D r}] - \sigma r e^{-m_D r} \, ,
\label{eq:vkmsre}
\ea
where $m_D^2 = 4 \pi N_c (1 + N_f/6) \alpha_s T^2/3$ is the in-medium gluonic Debye mass.  
Although there are arguments to support the use of the internal energy in thermally equilibrated systems \cite{PhysRevD.70.054507,PhysRevC.70.021901,SHURYAK200564}, it is unclear what the correct prescription is in the non-equilibrium case.  For this reason, one can consider the chosen real-part of the potential as a model choice.  To match smoothly onto the zero temperature limit we use
\be
\Re[V(r)] =
\begin{cases}  V_{\rm KMS}(r)  &\mbox{if } V_{\rm KMS}(r)  \leq V_{\rm vac}(r_{\rm SB}) \\
V_{\rm vac}(r_{\rm SB}) & \mbox{if } V_{\rm KMS}(r) > V_{\rm vac}(r_{\rm SB})
\end{cases} \, .
\label{eq:vmedre}
\ee
In the limit that $T\rightarrow0$, Eq.~\eqref{eq:vmedre} reduces to Eq.~\eqref{eq:vvac}.

The imaginary part of the potential is taken from a leading-order resummed perturbative QCD calculation of Laine et al 
\be
\Im[V(r)] = - C_F \alpha_s T \phi(m_D r) \, ,
\ee
with $\phi(\hat{r}) \equiv 1 - 2 \int_0^\infty \sin(z)/(z^2 + \hat{r}^2)^2 $ \cite{Laine:2006ns}.  We evaluate the strong coupling $\alpha_s$ at the scale $\mu = 2 \pi T$ and use three-loop running \cite{pdg} with \mbox{$\Lambda_{\overline{MS}} = 344$ MeV}, which reproduces the lattice result for the running coupling $\alpha_s(5\text{ GeV}) = 0.2034$~\cite{McNeile:2010ji}.

Using this complex potential, we then numerically solve the time-dependent Schr\"odinger equation on a discrete lattice.  The method used is manifestly unitary for real-valued potentials and is based on a split-step pseudospectral method \cite{Fornberg:1978,TAHA1984203}.  This algorithm allows for higher code accuracy and speed compared to traditional finite-size difference methods such as the Crank-Nicolson method and can be easily implemented on massively parallel architectures such as graphics cards~\cite{Boyd:2019arx}.  Due to the central nature of the potential, for a fixed orbital angular momentum $\ell$, we can reduce the problem to solving a one-dimensional Schr\"odinger equation for the scaled wave-function $u_\ell(r) = r \psi_\ell(r)$.  For the results reported herein, we used $N = 4096$ points with $L = r_{\rm max} = 19.7$ fm, resulting in a lattice spacing of $a \simeq 0.0048$ fm.  We compute the in-medium suppression for $\ell=0$ and $\ell=1$ states, separately.  We follow only the singlet evolution, however, singlet-octet transitions are in large part taken into account via the imaginary part of the singlet potential.  We will return to this issue in the conclusions where we discuss how to go beyond the description of the ensemble-averaged wave-function to include real-time transitions.  Finally, herein we do not consider $p_T$-dependence of the potential.  
The potential can depend on $p_T$ due to the fact that the quantum wave-packets are moving relative to the medium, however, the effect is expected to be sub-leading \cite{Escobedo:2011ie}.

\subsection{Initial condition for the quantum wave packets}

Due to the local nature of heavy quarkonium production, one can assume that the initial quantum mechanical wave-function is given by a Dirac delta function.  Since herein, we discretize space on a finite lattice, one must regulate the delta function.\footnote{One also expects the delta function to be physically smeared to a region on the size of $\sim 1/M_q$ for finite heavy quark masses.}  For this purpose, we choose a Gaussian initial wave-function
\be
u_\ell(r,\tau=0) \propto r^{\ell+1} \exp(-r^2/\Delta^2) \, ,
\ee
with $\Delta = 0.04$ fm.  For a given $\ell$, such an initial state is a quantum superposition of many eigenstates of the Schr\"odinger equation. After evolving the wave-function forward in time, the probability to find a given vacuum state can be obtained by computing the overlap of the in-medium quantum wave-function with the vacuum basis states.  In this way, one can obtain the survival probability of each state.  Due to the fact that the Hamiltonian for this system is non-Hermitian, one finds that these overlaps decay in time, which physically reflects the in-medium breakup of bottomonium states.  

Note that this is different than what has been done in prior works which compute bottomonium suppression using real-time solutions to the Schr\"odinger-Langevin equation \cite{Katz:2015qja,Gossiaux:2016htk,Bernard:2016spw}, since here the noise is encoded in the imaginary part of the potential, we use a realistic 3+1D hydrodynamics background tuned to data, and we solve the full 3+1D Schr\"odinger equation for ensembles of trajectories.  Finally, we note that one could also include non-trivial phases on each of the modes in the initial condition which could affect the evolution \cite{Bernard:2016spw}.  In this first work, we do not consider this possibility, however, it would be interesting to consider the role of initial phases in the future.

\subsection{Sampling initial production}

Since each wave-packet propagating through the QGP experiences a different temperature along its trajectory, we numerically solve the time-dependent Schr\"odinger equation for a large set of bottomonium trajectories (1.2 million).  Initial bottomonium production is Monte-Carlo sampled, assuming that the initial transverse spatial distribution is proportional to the binary overlap profile of the two colliding nuclei, $N_{AA}^\text{bin}(x,y)$.  We exploit the approximate boost-invariance of the QGP at mid-centrality and assume that all bottomonia have zero rapidity, $y=0$.  For the transverse momentum distribution, we assume that all states have a $p_T$-distribution proportional to $p_T/(p_T^2 + \langle M \rangle^2)^2$, where $\langle M \rangle$ is the average mass of all states being considered.  We assume that the initial azimuthal angle $\phi$ is distributed uniformly between 0 and $2\pi$.  

Once the initial position, momentum, and azimuthal angle are sampled, we then record the QGP temperature along the trajectory followed by the quantum wave-packets.  Herein, we assume that each quantum wave-packet's velocity is constant and, hence, they propagate along a straight line trajectory.  This would be exact in the limit of large quark masses.  In the future, one can also take into account the effect of in-medium scatterings and energy loss on the quantum wave-packets in order to relax this assumption.  Finally, we note that assuming that all cross-sections have the same $p_T$ dependence is an approximation.  In the future it would be interesting to investigate the role of different $p_T$ dependence of the production cross sections on our results \cite{Lansberg:2019adr}.

\subsection{Hydrodynamic background}

For the background temperature evolution, we use the output of a 3+1D quasiparticle anisotropic hydrodynamics (aHydroQP) code which has been tuned to reproduce soft hadron multiplicities, elliptic flow, etc at $\sqrt{s_{NN}} = 5.02$ TeV \mbox{\cite{mubarakforth,Alqahtani:2017mhy}}.  We use smooth optical Glauber initial conditions and the parameters used for the aHydroQP runs correspond to an initial central temperature of \mbox{$T_0 = 630$ MeV} at \mbox{$\tau_0 = 0.25$ fm/c}, with a constant specific shear viscosity of \mbox{$4\pi\eta/s = 2$}~\cite{mubarakforth}.  For each trajectory sampled, we evolve the quantum state using the in-medium complex potential until the local temperature is below the QGP transition temperature, \mbox{$T_\text{QGP} = 155$ MeV}.  We evolve the quantum wave-packets using the vacuum potential starting at $\tau = 0$ fm/c and turn on the in-medium potential at $\tau = \tau_\text{med} = 0.4$ fm/c.   In practice, we solve for the evolution in lab coordinates parameterized by Milne coordinates in the forward light cone and the time obtained from the hydrodynamics code is used as the evolution time for the states. Finally, whenever the temperature on a given trajectory drops below $T_\text{QGP}$, we use the vacuum potential for its evolution.  As a result, when $T<T_\text{QGP}$ the overlaps no longer evolve in time since (a) the potential becomes real and time-independent and (b) we use the numerically determined vacuum eigenstates of the vacuum potential which can be obtained by direct diagonalization of the problem \mbox{\eqref{eq:vvac}}.

\vspace{7mm}

\subsection{Excited state feed down}

After each quantum state is propagated along its trajectory, we convert the survival probabilities into particle number by multiplying by (1) the expected number of binary collisions in the centrality bin sampled and (2) the primordial production cross section for each bottomonium state.  In order for final state feed-down to result in the experimental observed $pp \rightarrow$ bottomonium production cross sections $\sigma_\text{exp} = \{ 57.6, 19, 13.82, 3.36, 2.07\}$ nb \cite{Khachatryan:2016xxp,CMS5TeV,Sirunyan:2018nsz,pwaveexp}, we take the primordial $pp \rightarrow \text{bottomonium}$ cross sections to be $\sigma_\text{primordial} = \{47.45, 24.95, 16.92, 4.057, 2.477\}$ nb, for the $\Upsilon(1s)$, $\Upsilon(2s)$, $\chi_b(1p)$, $\Upsilon(3s)$, and $\chi_b(2p)$ states, respectively. 

\begin{figure}[t]
\begin{center}
\includegraphics[width=0.975\linewidth]{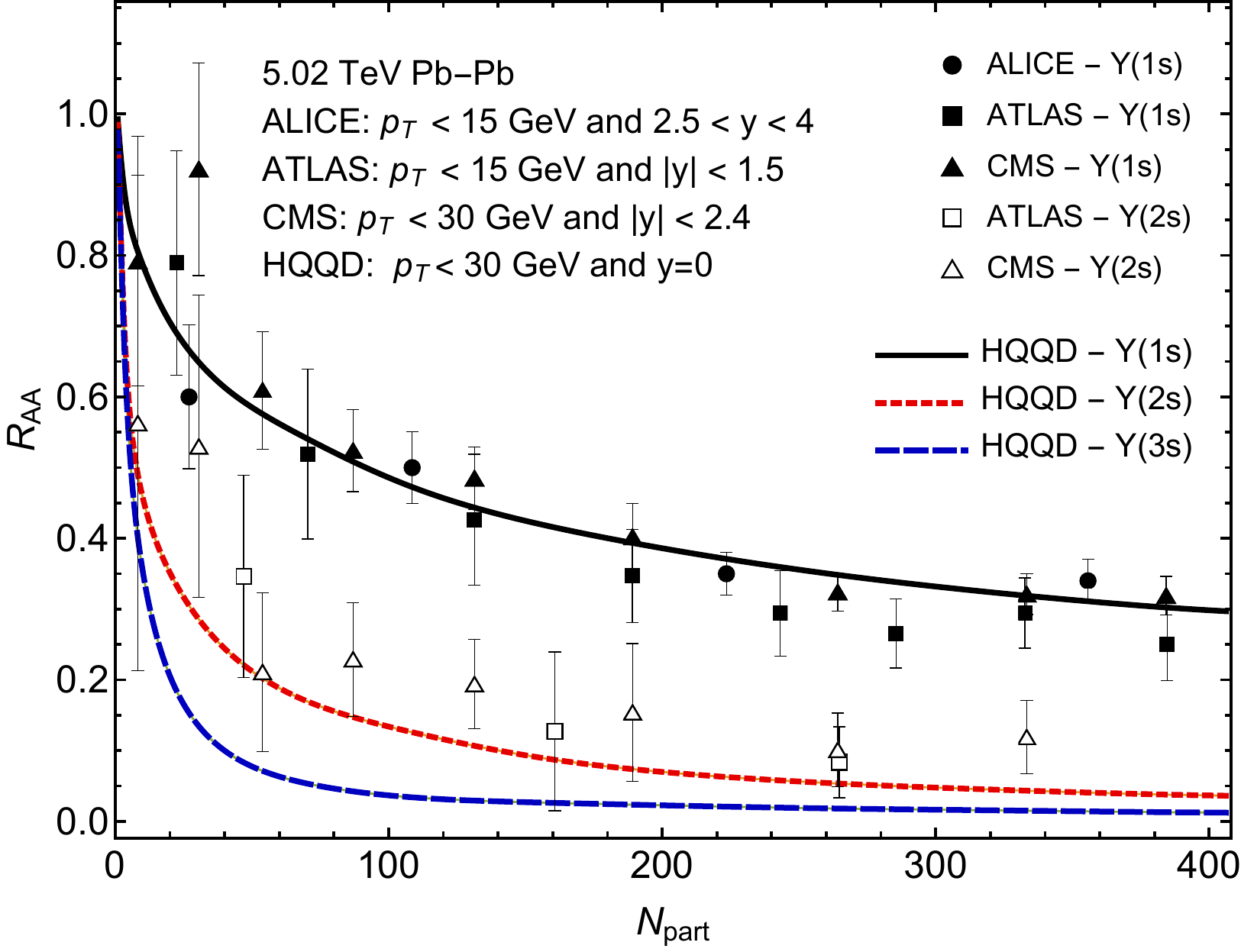}
\end{center}
\caption{Nuclear suppression factor, $R_{AA}$, of bottomonium $s$-wave states as a function of $N_\text{part}$.  The solid, short-dashed, and dashed lines show the predictions of HQQD.  Data points are from the  ALICE \cite{Acharya:2018mni}, ATLAS \cite{ATLAS5TeV}, and CMS \cite{Sirunyan:2018nsz} collaborations.  Experimental error bars shown were obtained by adding statistical and systematic uncertainties in quadrature.}
\label{fig:raavsnpart} 
\end{figure}

\begin{figure}[t]
\begin{center}
\includegraphics[width=0.975\linewidth]{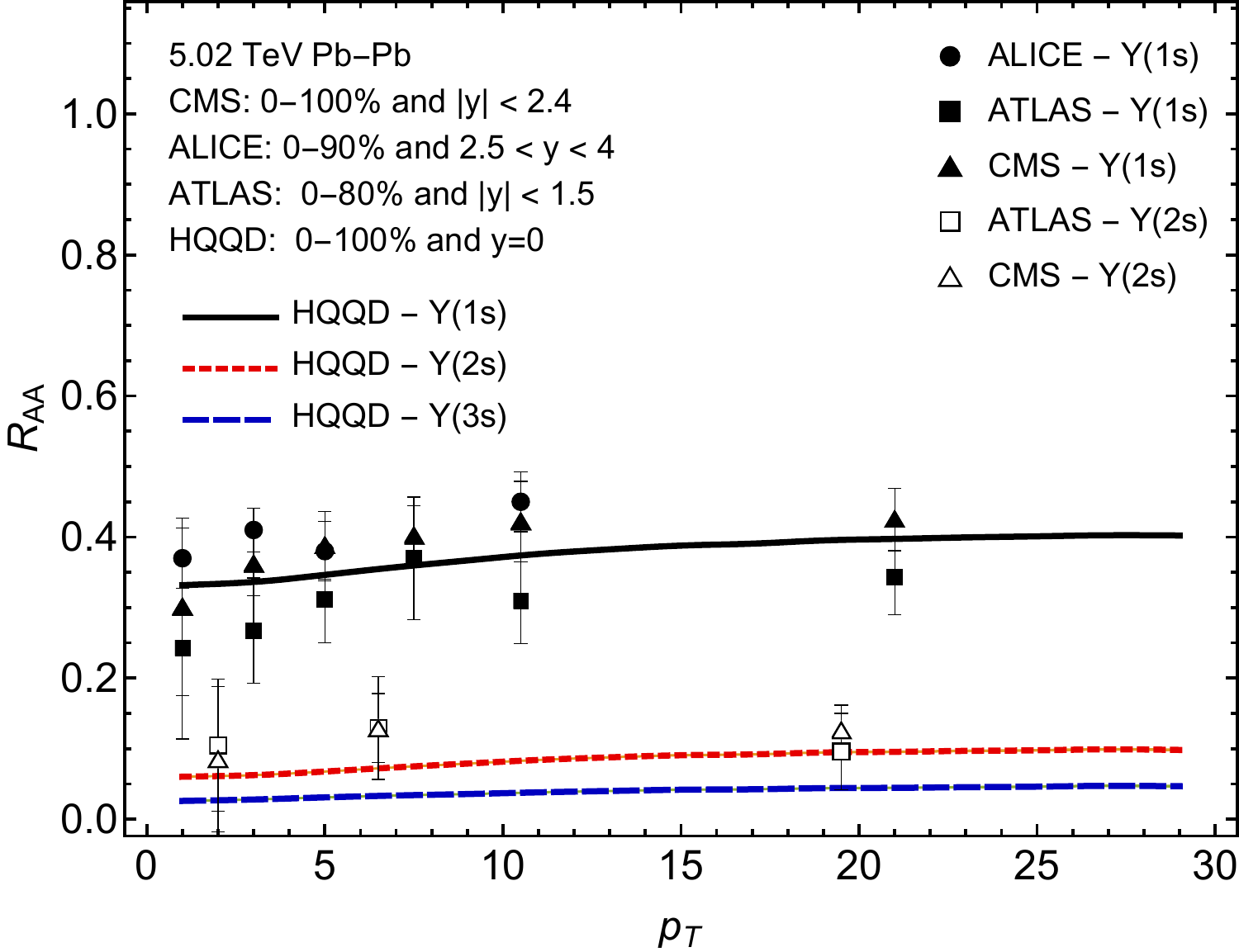}
\end{center}
\caption{Nuclear suppression factor, $R_{AA}$, of bottomonium $s$-wave states as a function of $p_T$.  Data sources used are the same as in Fig.~\ref{fig:raavsnpart}.}
\label{fig:raavspt} 
\end{figure}

To then account for final-state feed-down, we construct a vector $\vec{N}_\text{QGP}$ containing the numbers of each state produced and multiply it by a feed-down matrix, i.e. $\vec{N}_\text{final} = F \vec{N}_\text{QGP}$, with
\be
F = \left(
\begin{array}{cccccc}
 1 & 0.265 & 0.184 & 0.0657 & 0.0650  \\
 0 & 0.735 & 0 & 0.1060 & 0.0946  \\
 0 & 0 & 0.816 & 0 & 0.0047  \\
 0 & 0 & 0 & 0.8283 & 0  \\
 0 & 0 & 0 & 0 & 0.8357 \\
\end{array}
\right) .
\ee
This matrix is constructed from the experimentally measured branching ratios of the various bottomonium states \cite{pdg}.\footnote{In the case of states with hyperfine splitting, e.g. $\chi_{b0}$. $\chi_{b1}$, and $\chi_{b2}$, we have averaged the branching ratios.}  For example, the final number of 1s states produced can be computed as $N^{\Upsilon(1s)}_\text{final} = f_{11} N^{\Upsilon(1s)}_\text{QGP} + f_{12} N^{\Upsilon(2s)}_\text{QGP} +  f_{13} N^{\chi_b(1p)}_\text{QGP} + f_{14} N^{\Upsilon(3s)}_\text{QGP} +  f_{15} N^{\chi_b(2p)}_\text{QGP}$.  Note that each column of $F$ must sum to unity in order to preserve bottom number.  

\subsection{Computation of $R_{AA}$ and $v_n$ using HQQD}

To compute $R_{AA}$, we divide the final number of bottomonium states produced by the number of binary collisions in the sampled centrality class times the post feed-down $pp$ production cross-section for each state.  Since we know the reaction plane (provided by aHydroQP) one has $\Psi_\text{RP} = 0$ and, as a result, one can compute $v_n$ by simply averaging $\cos(n\phi)$ over all particles, $v_n \equiv \langle \cos(n\phi) \rangle$, in a given $p_T$ and centrality bin.  For both $R_{AA}$ and $v_2$, we report the statistical uncertainty associated with the average over the sampled quantum wave-packet trajectories.  For both observables, we ignore any possible cold nuclear matter effects, however, these can be included in the future.  The resulting model will be referred to as Heavy Quarkonium Quantum Dynamics (HQQD) in what follows.

\section{Results}  

\begin{figure}[t]
\begin{center}
\includegraphics[width=0.975\linewidth]{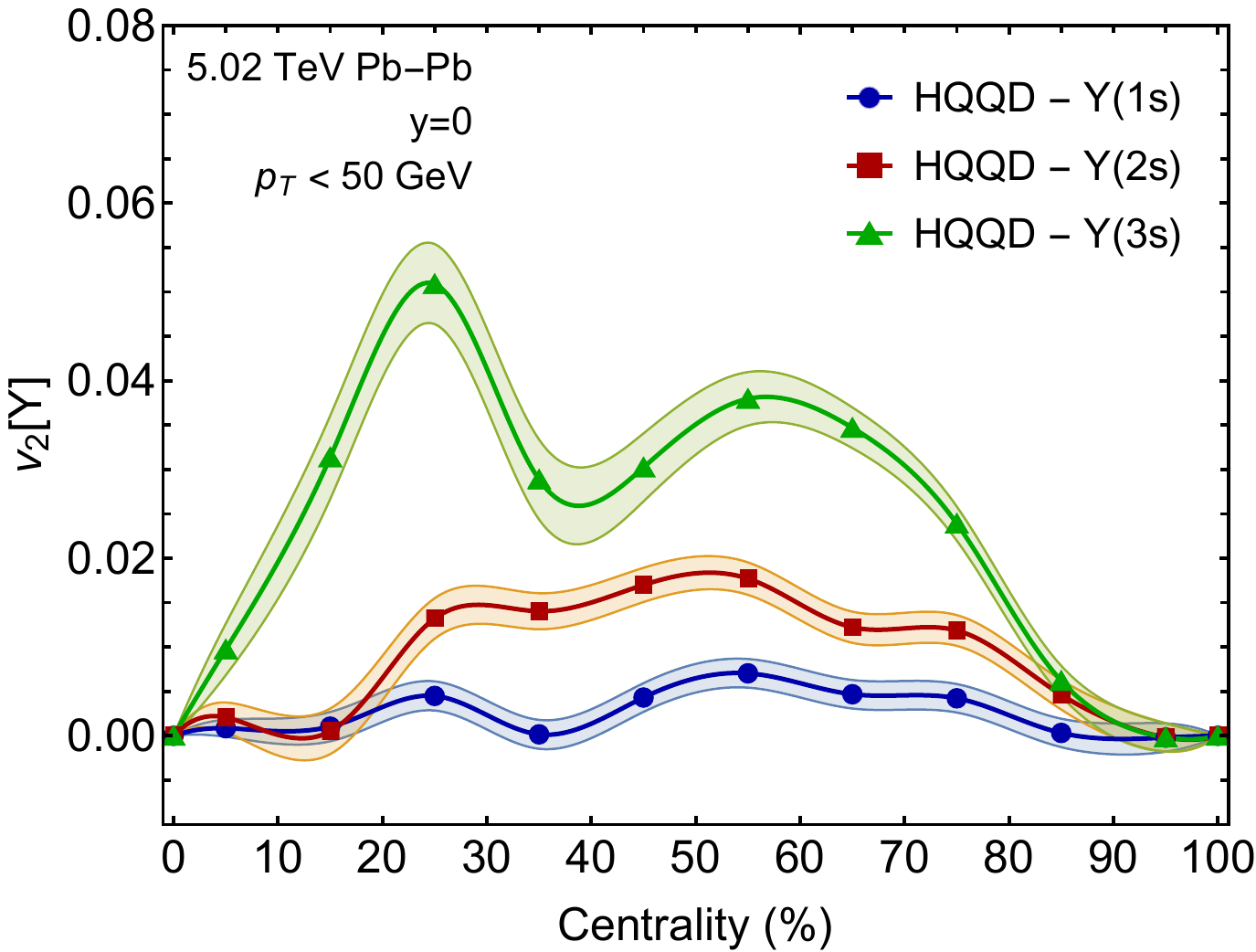}
\end{center}
\caption{Elliptic flow for $s$-wave bottomonium states as a function of centrality. Solid lines and bands show spline-interpolated results for the mean and statistical uncertainty of the mean obtained from HQQD.  Points show results obtained in equally spaced bins of 10\% centrality from 0-100\%.} 
\label{fig:v2_swave}
\end{figure}

\begin{figure}[t]
\begin{center}
\includegraphics[width=0.975\linewidth]{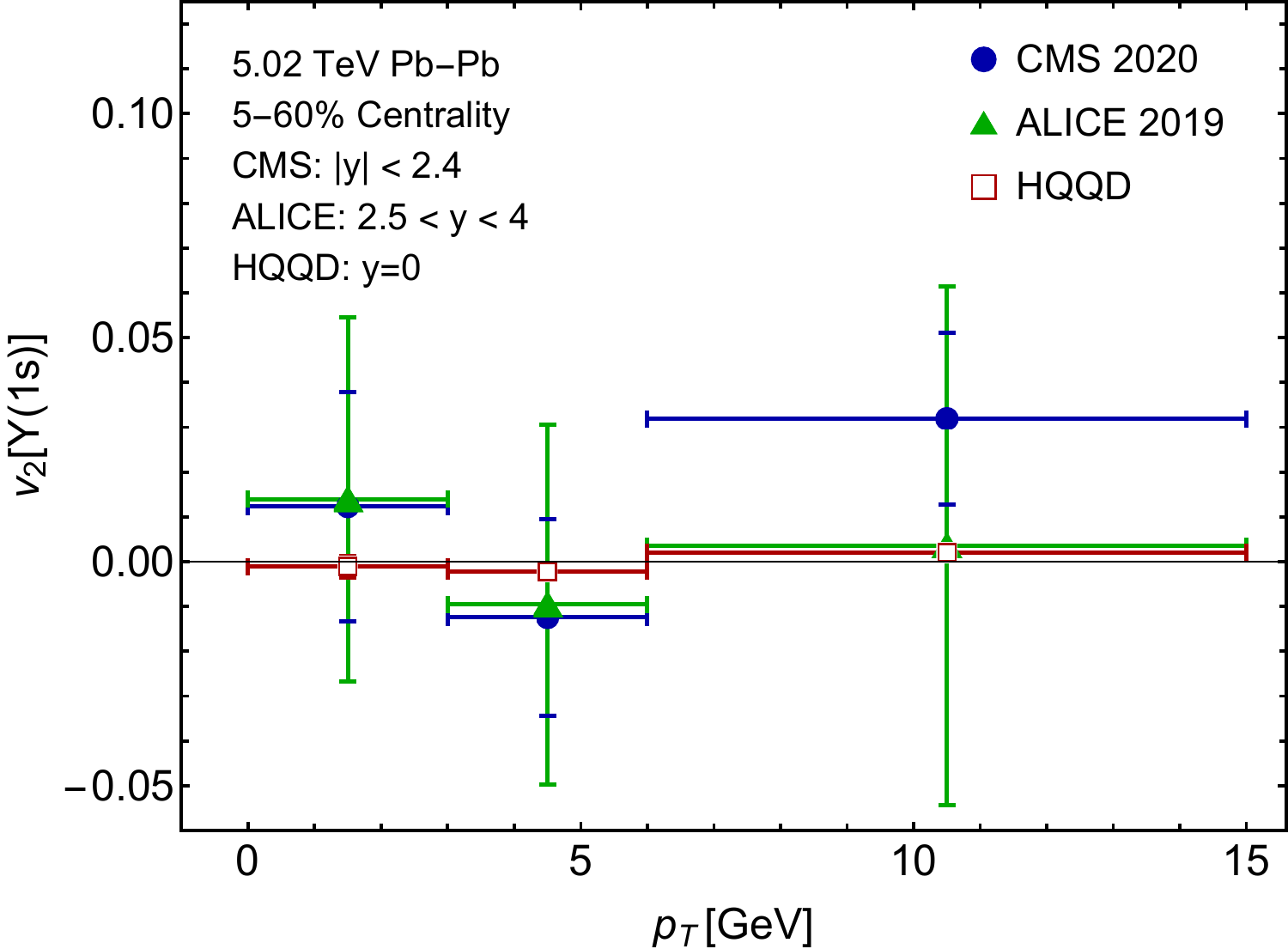}
\end{center}
\caption{The elliptic flow $v_2[\Upsilon(1s)]$ as a function of $p_T$ in three $p_T$-bins.  Open red squares are HQQD predictions and the data are from the ALICE \cite{Acharya:2019hlv} and CMS \cite{Sirunyan:2020qec} collaborations. } 
\label{fig:v21s}
\end{figure}

In Fig.~\ref{fig:raavsnpart} we present HQQD predictions for the suppression of $\Upsilon(1s)$, $\Upsilon(2s)$, and $\Upsilon(3s)$ states as a function of $N_{\rm part}$.  For this Figure, in HQQD we applied a transverse momentum cut of $p_T < 30$ GeV.  We compare with results obtained by the ALICE \cite{Acharya:2018mni}, ATLAS \cite{ATLAS5TeV}, and CMS \cite{Sirunyan:2018nsz} collaborations, shown as circles, squares, and triangles, respectively.  From this Figure, we see that HQQD does a quite reasonable job in describing the $N_\text{part}$ dependence of $R_{AA}[\Upsilon(1s)]$, however, HQQD predicts a somewhat smaller $R_{AA}[\Upsilon(2s)]$ than the experimental results.  Similar conclusions can be obtained from Fig.~\ref{fig:raavspt}, where we present  $R_{AA}[\Upsilon]$ as a function of transverse momentum.  For this Figure, we averaged over centrality with a weight function $w(c) = \exp(-c/20)$, with $c \in [0,100]$. This weight function reflects the experimentally observed distribution of the number of $\Upsilon$ states versus centrality~\cite{Chatrchyan:2012np}. From the results shown in Fig.~\ref{fig:raavspt}, we see that HQQD predicts a very weak dependence of $R_{AA}[\Upsilon]$ on $p_T$, with only a small decrease at momentum less than the mass scale of the bottomonium states.  The increased suppression at low-$p_T$ can be attributed to such wave-packets having, on average, a longer effective lifetime inside the QGP fireball (due to their lower velocities).

In Fig.~\ref{fig:v2_swave}, we present our results for the elliptic flow of $\Upsilon(1s)$, $\Upsilon(2s)$, and $\Upsilon(3s)$ states as a function of centrality.  For this Figure, we impose $p_T < 50$ GeV and compute $v_2$ in 10 equally spaced centrality bins from 0-100\%.  The bands in this Figure show the statistical uncertainty associated with the mean values extracted in each bin.  As can be seen from this Figure, there is a clear ordering of the elliptic flow, with the $\Upsilon(3s)$ state having the largest flow and the $\Upsilon(1s)$ the smallest.  This is in agreement with expectations, since the source of the elliptic flow in all cases is the suppression of the states and, hence those with stronger suppression will have a larger elliptic flow.  One other thing that is evident from Fig.~\ref{fig:v2_swave} is that the elliptic flow of all states goes to zero for central collisions (left hand side of the plot).  This, of course, is a consequence of our choice of non-fluctuating optical Glauber initial conditions and provides a non-trivial test of the HQQD calculation of $v_2$. If one includes geometric fluctuations in the initial hydrodynamic variables (energy density, etc.), one would expect to see small, but finite, values for the elliptic flow of all states in central collisions.  On the right hand side of Fig.~\ref{fig:v2_swave} one sees that the elliptic flow for all states goes to zero.  This, again, agrees with expectations since the QGP lifetime in such ultraperipheral events is zero.

One other feature visible in Fig.~\ref{fig:v2_swave} are the oscillations in the elliptic flow vs centrality for all states shown.  For the $\Upsilon(3s)$, there is a very clear oscillation visible.  In the HQQD calculation, these oscillations are caused by quantum mechanical oscillations in the state overlaps due to the time-dependent potential.  The characteristic period of these oscillations is on the order of a fm/c and, due to differences in the average path length traversed in each centrality bin, can result in oscillations in $v_2$.  We note that $v_2 = \langle \cos(2\phi) \rangle$ is more sensitive to these oscillations because it explicitly involves, for example, differences between the survival probability along the short and long sides of the QGP.  In $R_{AA}$ one averages over all angles with the same weight and, as a result, these oscillations are smoothed out in the average over quantum wave-packet trajectories.  On the contrary, since for $v_2$ different angles contribute with different weights/signs, it is naturally more sensitive to small differences in the survival probability and hence is more sensitive to these oscillations.

\begin{figure}[h]
\begin{center}
\includegraphics[width=0.95\linewidth]{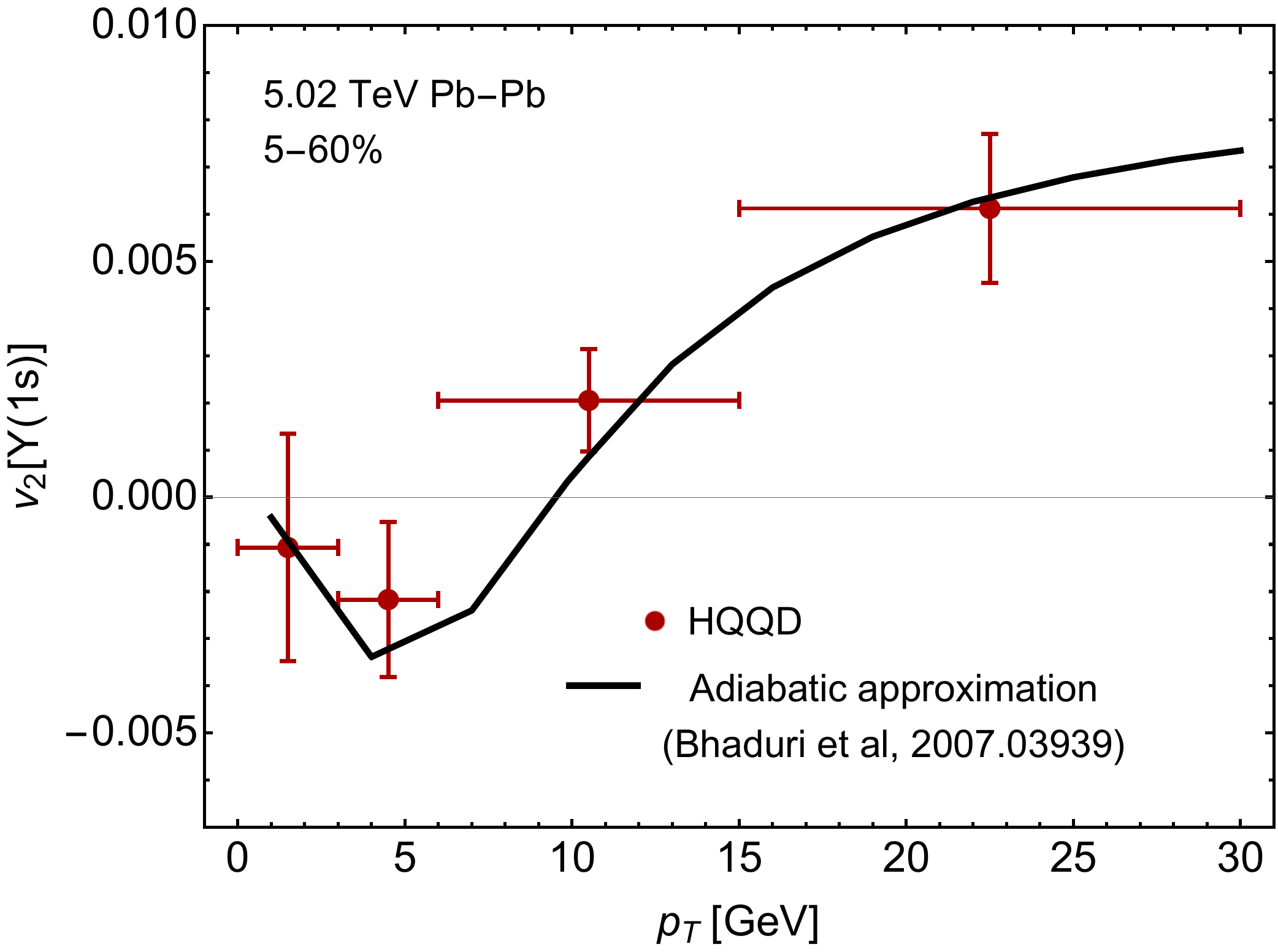}
\end{center}
\caption{
Comparison of the HQQD predictions for $v_2[\Upsilon(1s)]$ versus $p_T$ with prior results obtained using the adiabatic approximation \cite{Bhaduri:2020lur}.  The black solid line is the result from the adiabatic approximation and the red points are the HQQD results, with the horizontal error bar indicating the $p_T$-bin used to compute the result in HQQD and the vertical error bar indicating the statistical uncertainty associated with the average over trajectories in HQQD.
}
\label{fig:adiabatic-comparison}
\end{figure}

We turn next to Fig.~\ref{fig:v21s} in which we present a comparison of HQQD predictions for $v_2[\Upsilon(1s)]$ with experimental data collected by the ALICE \cite{Acharya:2019hlv} and CMS \cite{Sirunyan:2020qec} collaborations in three different transverse momentum bins: 0-4, 4-6, and 6-15 GeV.  For both HQQD and the experiments, the results are integrated over centrality in the range \mbox{5-60\%}.  As can be seen from this Figure, HQQD predicts a result consistent with zero in the lowest $p_T$ bin, a slightly negative result in the central bin, and a small but positive value in the highest momentum bin.  This trend (positive near zero, then negative, and then positive again) and the overall magnitude of $v_2$ predicted by HQQD  is similar to what has been predicted previously using a model which relies on the adiabatic approximation \cite{Bhaduri:2020lur}.  In Ref.~\cite{Bhaduri:2020lur} it was posited that the explanation for this negative $v_2$ is related to the transverse expansion of the QGP overtaking bottomonia states which have escaped from near the surface of the QGP.\footnote{See Fig.~3 of Ref.~\cite{Bhaduri:2020lur} and the surrounding discussion.} With respect to the comparisons with experimental data, we find reasonable agreement with available data, given current experimental uncertainties, and one sees a similar trend in the three centrality classes as predicted by HQQD.

\begin{figure}[t!]
\begin{center}
\includegraphics[width=0.975\linewidth]{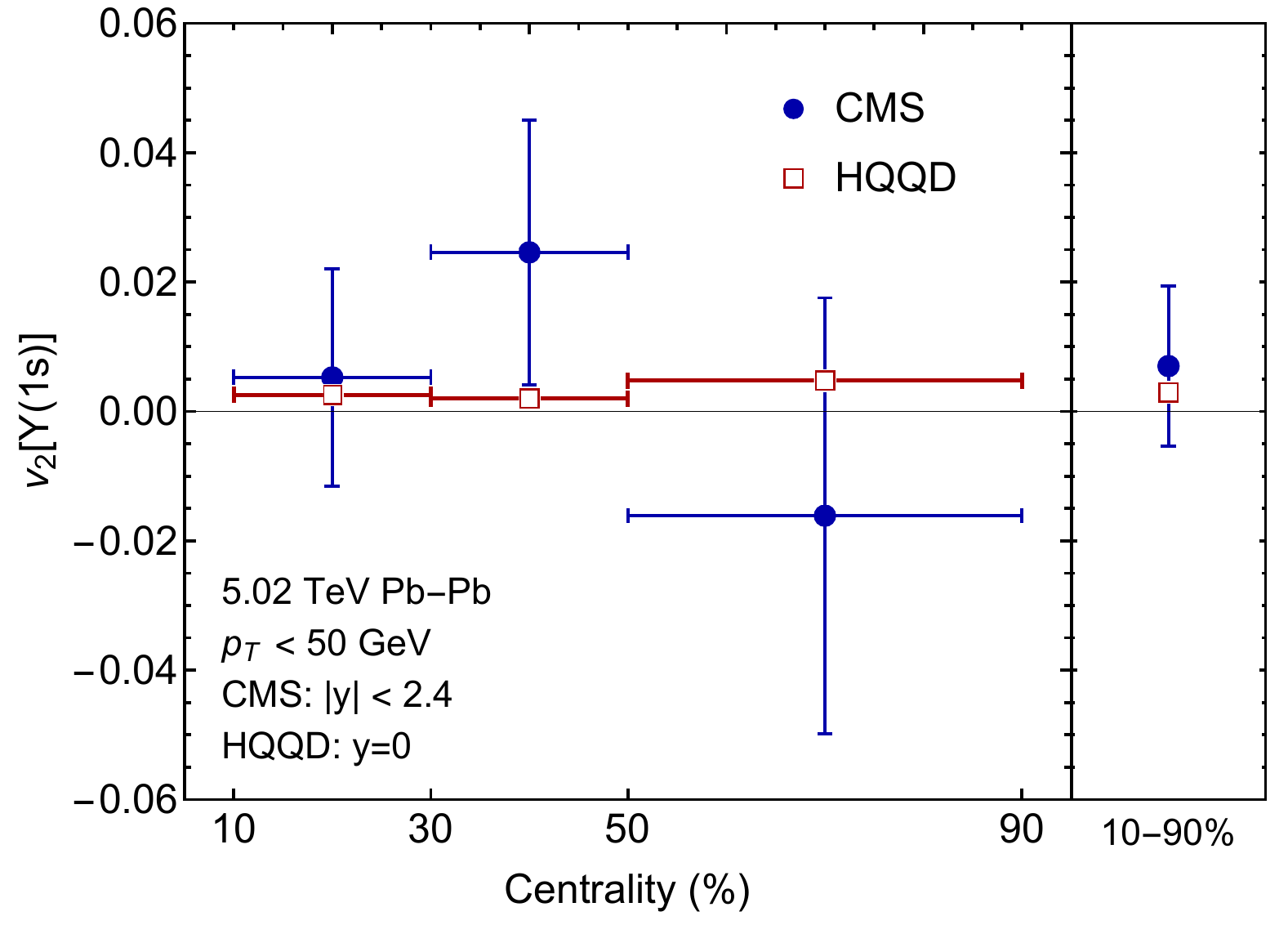}
\end{center}
\caption{Centrality dependence of $v_2[\Upsilon(1s)]$ shown in 10-30\%, 30-50\%, 50-90\%, and 10-90\% centrality bins. Open squares are predictions of HQQD.}
\label{fig:v21sCent} 
\end{figure}

It is possible to make quantitative comparisons between the predictions of HQQD and prior calculations of $v_2[\Upsilon(1s)]$ in order to assess whether or not $v_2$ is sensitive to the full quantum dynamics.  For this purpose, in Fig.~\ref{fig:adiabatic-comparison} we compare the $p_T$-dependence of  $v_2[\Upsilon(1s)]$ with the result reported in Ref.~\cite{Bhaduri:2020lur}.  In order to increase statistics, the HQQD results (red points) are binned in $p_T$.  The error bars listed for HQQD are computed using the statistical uncertainty associated with the average over trajectories.  For both results we integrate over centrality in the range 5-60\%.  As can be seen from Fig.~\ref{fig:adiabatic-comparison}, the two calculations are in very good agreement with one another for the 5-60\% centrality-integrated $v_2[\Upsilon(1s)]$.  Additionally, we stress that both calculations predict a negative $v_2$ with a maximum negative value around $p_T \sim 5$ GeV.  Note that the agreement between HQQD and results obtained using the adiabatic approximation for the $p_T$ dependence of $R_{AA}[\Upsilon(1s)]$ is also quite good at low transverse momentum, with the two models predicting approximately the same $R_{AA}[\Upsilon(1s)]$ at $p_T < 10$ GeV.  At higher $p_T$ the adiabatic approximation results for $R_{AA}[\Upsilon(1s)]$ (Figure 1 of Ref.~\cite{Bhaduri:2020lur}) indicate less suppression than found in HQQD.  This is most likely due to the manner in which finite formation time effects were included in the adiabatic approximation.

\begin{figure}[t]
\begin{center}
\includegraphics[width=0.975\linewidth]{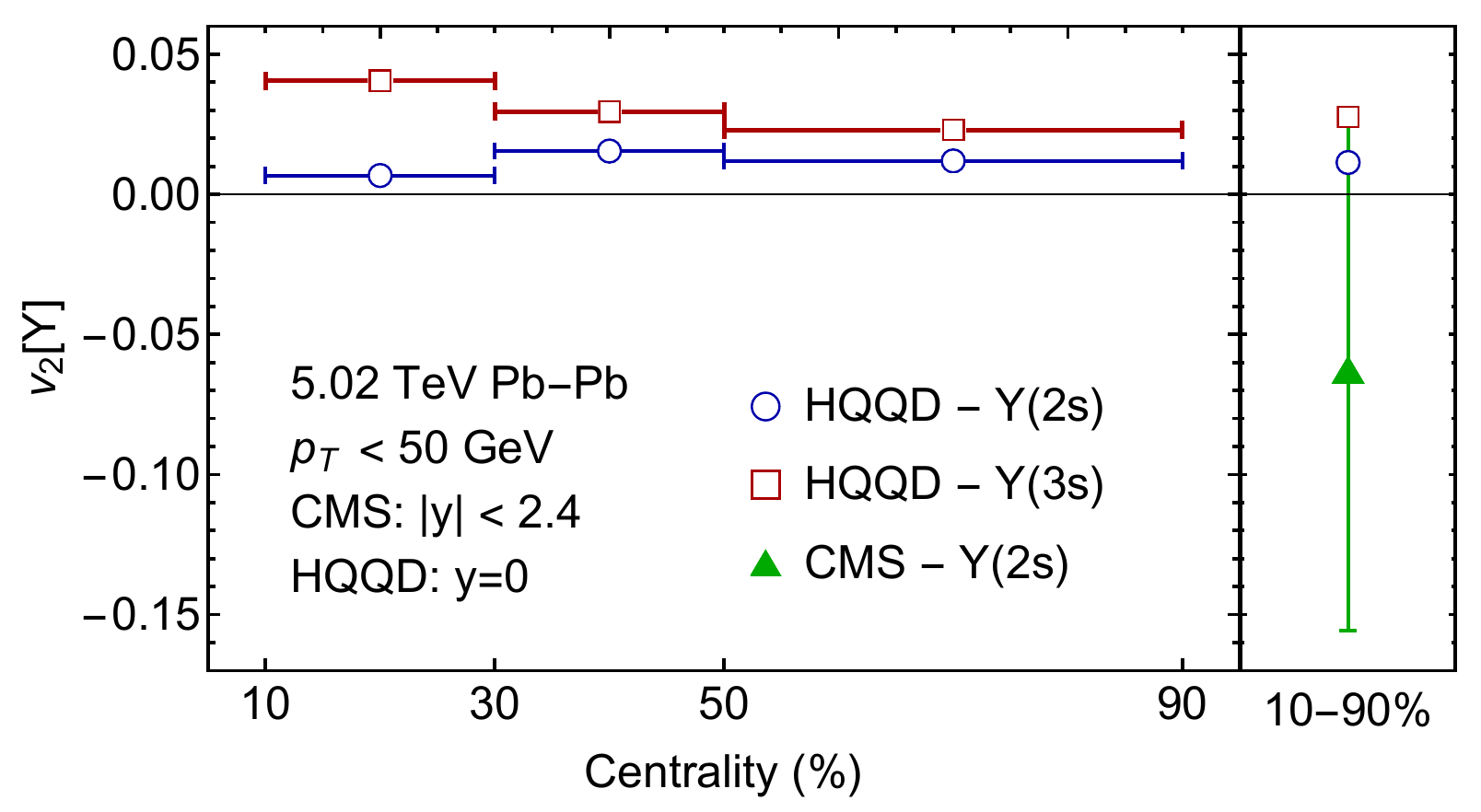}
\end{center}
\caption{Centrality dependence of $v_2[\Upsilon(2s)]$ and $v_2[\Upsilon(3s)]$ in the same centrality bins as Fig.~\ref{fig:v21sCent}.  Open symbols are predictions of HQQD.  In the 10-90\% class we include recent data reported by the CMS collaboration for integrated $v_2[\Upsilon(2s)]$ \cite{Sirunyan:2020qec}.} 
\label{fig:v2excitedCent} 
\end{figure}

In Fig.~\ref{fig:v21sCent}, we present a comparison of HQQD with experimental data from the CMS collaboration for the centrality dependence of $v_2[\Upsilon(1s)]$. All results are binned into three centrality bins:  10-30\%, 30-50\%, and 50-90\%.  In the rightmost panel of Fig.~\ref{fig:v21sCent}, we show the experimental result integrated over 10-90\% centrality compared to the HQQD prediction in the same centrality interval.  From this Figure we see that the integrated $v_2[\Upsilon(1s)]$ in the 10-90\% class is in agreement, within uncertainties, with the experimental data provided by CMS.  In the separate bins (left panel), we see good agreement in the 10-30\% bin, however, in the other two bins we larger differences, albeit still within $2\sigma$ of the HQQD predictions.  In the future, hopefully higher statistics will allow for more constraining comparisons between HQQD and experiment.  

In Fig.~\ref{fig:v2excitedCent}, we present HQQD predictions for $v_2[\Upsilon(2s)]$ and $v_2[\Upsilon(3s)]$ in the same centrality bins as Fig.~\ref{fig:v21sCent}.  For $v_2[\Upsilon(2s)]$, there is currently only one integrated data point available from the CMS collaboration, which is shown as a green triangle in the 10-90\% panel (right).  Comparing the integrated results, we see that  $v_2[\Upsilon(2s)]$ is currently within the reported experimental uncertainties, however, at the very top end of them.  Again, increased statistics will allow for more accurate comparisons in the future.  In the left panel of Fig.~\ref{fig:v2excitedCent} we see that the flow of  $v_2[\Upsilon(3s)]$ can be on the same order of magnitude as the experimentally observed $v_2[J/\psi]$~\cite{Acharya:2019hlv,Sirunyan:2020qec}.

\begin{figure}[t!]
\centerline{
\includegraphics[width=0.975\linewidth]{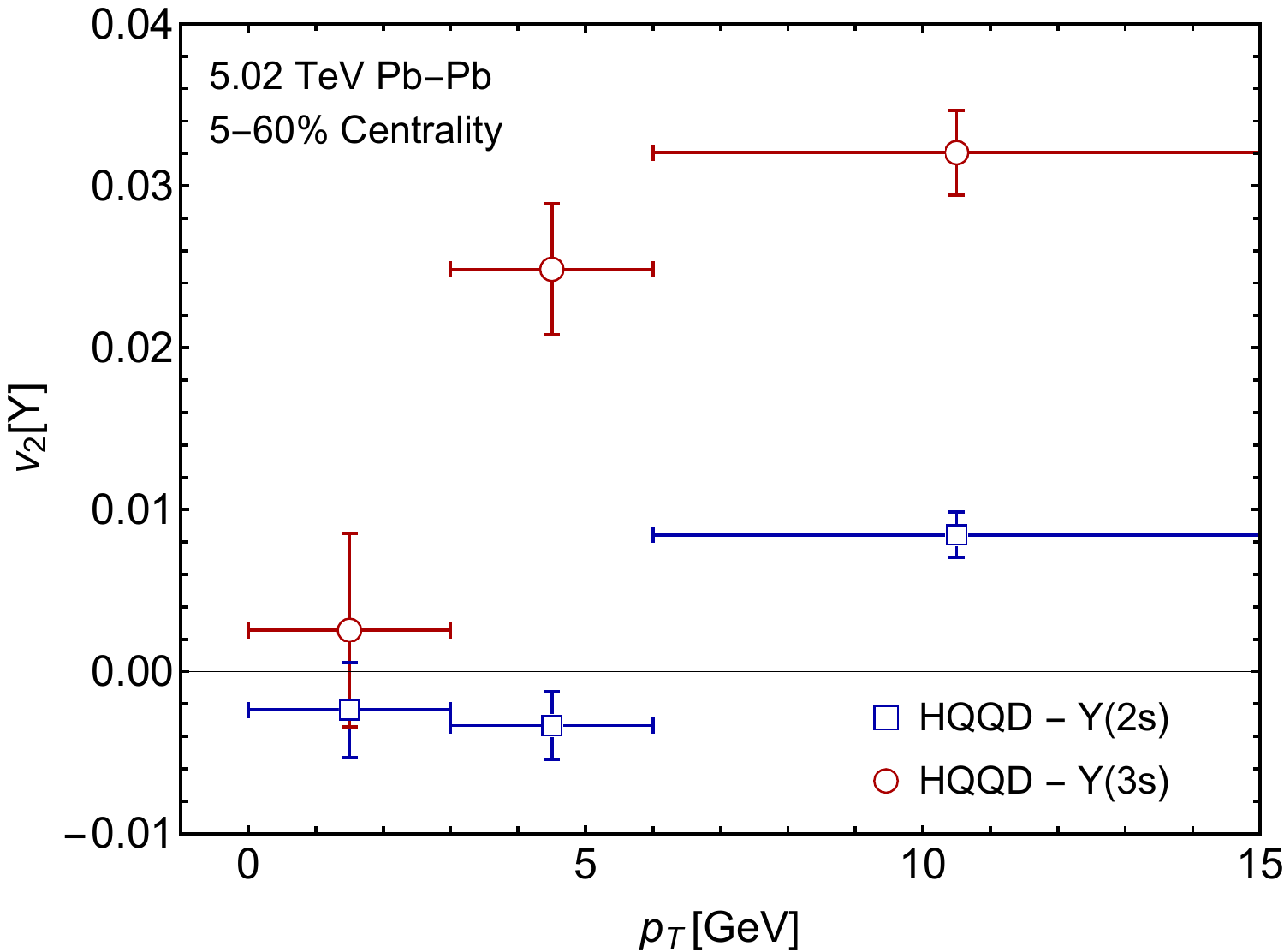}
}
\caption{HQQD predictions for the elliptic flow of $\Upsilon(2s)$ and $\Upsilon(3s)$ states in the 5-60\% centrality bin.  The transverse momentum bins are the same as used in Fig.~\ref{fig:v21s}.}
\label{fig:v22s3s}
\end{figure}

In Fig.~\ref{fig:v22s3s}, we present HQQD predictions for $v_2[\Upsilon(2s)]$ and $v_2[\Upsilon(3s)]$ as a function of transverse momentum using the same $p_T$-bins as Fig.~\ref{fig:v21s} in order to allow for easier comparison with experimental data in the future.  From this Figure we see that the $\Upsilon(3s)$ can develop a sizable $v_2$ solely due to path length differences between the short and long sides of the QGP fireball.  Turning to the $\Upsilon(2s)$ we see that, similar to the $\Upsilon(1s)$, HQQD predicts a negative $v_2$ in the lowest two $p_T$-bins.  This once again is related to the fact that the QGP expands more rapidly along the short side than the long side, which can have the affect of overtaking bottomonium states which had previously escaped the QGP with $\phi \sim 0$.   In the highest $p_T$-bin shown, we see that HQQD predicts positive $v_2$ for both states.

Finally, in Table \ref{tab:comp} we present comparisons between HQQD predictions for various observables and the corresponding experimental results from the ALICE, ATLAS, and CMS collaborations.  In this Table, the results are integrated over centrality and transverse momentum in the ranges shown in the middle column and the last row shows the HQQD prediction for $v_2[\Upsilon(3s)]$.  We do not indicate the rapidity cuts used by each experimental collaboration, which correspond to $2.5 < y < 4.0$, $|y| < 1.5$, and $|y| < 2.4$ for the ALICE, ATLAS, and CMS collaborations, respectively.  From this Table we see that all HQQD predictions are within the combined statistical and systematic uncertainties reported for each measurement.  We once again note that, for the $\Upsilon(2s)$, HQQD seems to predict slightly too much suppression, however, the HQQD predictions are still compatible with experimental results within uncertainties.  

\begin{table}[t!]
\begin{center}
\def\arraystretch{1.2}
{\scriptsize
\begin{tabular}{|c|c|c|c|}
\hline
{\bf ~Observable~} & {\bf ~Source/Cuts~}  &{\bf ~Experiment/HQQD~}  \\
\hline
$R_{AA}[\Upsilon(1s)]$ & ALICE 0-90\%  \cite{Acharya:2018mni}& 0.37 $\pm$ 0.02 $\pm$ 0.03  \\
& $p_T < 15$ GeV & 0.3556 $\pm$ 0.0002 \\
\hline
$R_{AA}[\Upsilon(1s)]$ & ATLAS 0-80\% \cite{ATLAS5TeV} & 0.32 $\pm$ 0.02 $\pm$ 0.05   \\
& $p_T < 30$ GeV & 0.3673 $\pm$ 0.0002 \\
\hline
$R_{AA}[\Upsilon(1s)]$ & CMS 0-100\% \cite{Sirunyan:2018nsz}& 0.376 $\pm$ 0.013 $\pm$ 0.035  \\
& $p_T < 30$ GeV & 0.3673 $\pm$ 0.0002 \\
\hline
$R_{AA}[\Upsilon(2s)]$ & ALICE 0-90\% \cite{Acharya:2018mni}& 0.10 $\pm$ 0.04 $\pm$ 0.02  \\
& $p_T < 15$ GeV & 0.07000 $\pm$ 0.00006 \\
\hline
$R_{AA}[\Upsilon(2s)]$ & ATLAS 0-80\% \cite{ATLAS5TeV} & 0.11 $\pm$ 0.04 $\pm$ 0.04   \\
& $p_T < 30$ GeV & 0.07920 $\pm$ 0.00006 \\
\hline
$R_{AA}[\Upsilon(2s)]$ & CMS 0-100\% \cite{Sirunyan:2018nsz}& 0.117 $\pm$ 0.022 $\pm$ 0.019  \\
& $p_T < 30$ GeV & 0.07920 $\pm$ 0.00006 \\
\hline
$R_{AA}[\Upsilon(3s)]$ & CMS 0-100\% \cite{Sirunyan:2018nsz}& 0.022 $\pm$ 0.038 $\pm$ 0.016   \\
& $p_T < 30$ GeV & 0.03622 $\pm$ 0.00004 \\
\hline
$v_2[\Upsilon(1s)]$ & ALICE 5-60\% \cite{Acharya:2019hlv}& -0.003 $\pm$ 0.030 $\pm$  0.006  \\
& $2 < p_T < 15$ GeV & 0.0006 $\pm$ 0.0009 \\
\hline
$v_2[\Upsilon(1s)]$ & CMS 10-90\% \cite{Sirunyan:2020qec}& 0.007 $\pm$ 0.011 $\pm$ 0.005 \\
 & $p_T < 30$ GeV & 0.0026 $\pm$ 0.0007 \\
 \hline
$v_2[\Upsilon(2s)]$ & CMS 10-90\% \cite{Sirunyan:2020qec}& -0.063 $\pm$ 0.085 $\pm$ 0.037  \\
 & $p_T < 30$ GeV & 0.0105 $\pm$  0.0008 \\
 \hline
 $v_2[\Upsilon(3s)]$ & HQQD 10-90\% & N/A \\
 & $p_T < 30$ GeV & 0.0264 $\pm$ 0.0011 \\
 \hline
\end{tabular}
}
\end{center}
\caption{Comparison of HQQD predictions for integrated $R_{AA}[\Upsilon]$ and $v_2[\Upsilon]$ with available experimental data.  The first column indicates the observable, the second column indicates the source of the experimental result and relevant cuts, and the third column shows the experimental result on the first line and the HQQD prediction on the second line. For all experimental results, the first uncertainty reported is statistical uncertainty and the second is systematic uncertainty.  For HQQD, the uncertainties reported are statistical uncertainties associated with the average over trajectories.}
\label{tab:comp}
\end{table}

\section{Conclusions and outlook} 

In this paper we used real-time quantum evolution to compute the suppression and elliptic flow of bottomonium states and presented the details of the resulting HQQD model.  Using HQQD, we sampled a large set of bottomonium trajectories (1.2 million).  For the HQQD hydrodynamic background, we used anisotropic hydrodynamics to provide the 3+1D temperature field through which the states were propagated.  Given this background, we then solved the time-dependent Schr\"{o}dinger equation with a complex potential and obtained the survival probability for each bottomonium state by computing the quantum mechanical overlap of the in-medium evolved wave-packet with the vacuum eigenstate of the state of interest.  

After averaging over all wave-packet trajectories, we were able to obtain precise estimates for $R_{AA}$ which are in quite reasonable agreement with available experimental data.  We then presented predictions of HQQD for the elliptic flow of $\Upsilon(1s)$,  $\Upsilon(2s)$, and  $\Upsilon(3s)$ and compared those to available experimental data.  In the case of $v_2$ we, once again, found reasonable agreement between theory and experiment, with the integrated $v_2$ being consistent with the experimental data within uncertainties (Table \ref{tab:comp}).  With respect to $v_2$, we emphasized two model observations:  (1) that $v_2$ for the various states can be non-monotonic (oscillating) due to quantum mechanical oscillations in the time-evolved overlaps and (2) that $v_2$ for $\Upsilon(1s)$ and $\Upsilon(2s)$ can be negative in intermediate transverse momentum bins, e.g. $4 < p_T < 6$ GeV.  The first observation is novel, however, the second has been observed previously in calculations of $v_2$ using the adiabatic approximation \cite{Bhaduri:2020lur}.  In order to make the HQQD predictions presented herein, we had to make a set of model choices corresponding to, for example, the choice of the real and imaginary parts of the quark-antiquark potential and the precise initialization time for medium interaction $\tau_\text{med}$.  We did not attempt to estimate the systematic uncertainties associated with these choices, but plan to in a forthcoming paper.

Our use of real-time solutions allowed us to go beyond the adiabatic approximation.  Overall, we found our HQQD results to be qualitatively consistent with previous adiabatic approximation results, however, with HQQD one has a more complete description of the quantum dynamics.  The use of real-time solutions allowed us, for example, to include the effect of quantum-mechanical state mixing due to the time-dependent in-medium quark-antiquark potential.  Looking to the future, we plan to more fully include the effect of thermal noise in the underlying evolution.  In this work, we evolved the states with a complex Hamiltonian which is appropriate for describing the evolution of the average wave-function of the system.  As a result, the system remains in the singlet configuration and there can be no transitions between different angular momentum states.

The description in terms of the average wave-function is not yet a complete description, however, in practice one finds that the in-medium wave-function evolution for the ground state is well-approximated by the evolution of the average wave-function subject to a complex Hamiltonian when including noisy potentials \cite{munichforth,alexanderforth}.  For excited state suppression, it may be important to go beyond the complex Hamiltonian approach used herein.  For this purpose, one can either introduce a noisy potential, with the noise spectrum set by the imaginary part of the quark-antiquark potential \cite{Akamatsu:2011se,Akamatsu:2012vt,Rothkopf:2013ria,Rothkopf:2013kya,Kajimoto:2017rel,Akamatsu:2018xim,Miura:2019ssi,Sharma:2019xum} or instead solve the resulting Lindblad equation including both singlet and octet states in order to describe the evolution of the full density matrix \cite{Akamatsu:2014qsa,Brambilla:2016wgg,Brambilla:2017zei,Blaizot:2017ypk,Blaizot:2018oev}.  Preliminary results obtained using the quantum trajectories method to solve the Lindblad equation indicate that the more accurate inclusion of noise effects and singlet-octet transitions will result in only small changes in $R_{AA}[\Upsilon(1s)]$, however, these studies indicate that the excited states are less suppressed when including initial state octet production and in-medium stochastic singlet-octet transitions~\cite{munichforth}.  

This will hopefully result in better agreement between HQQD and experimental results for $R_{AA}[\Upsilon(2s)]$.  Finally, we mention again that it would also be interesting to study the effect of geometric fluctuations in the initial state on $v_2[\Upsilon]$.

\section*{Acknowledgments}

We thank the participants of the EMMI Rapid Reaction Task Force meeting on ``Suppression and (re)generation of quarkonium in heavy-ion collisions at the LHC'' for useful discussions and are grateful to N. Brambilla for providing feedback on an early version of this manuscript.  We also thank the Ohio Supercomputer Center under the auspices of Project No.~PGS0253.  M.S. and A.I. were supported by the U.S. Department of Energy, Office of Science, Office of Nuclear Physics Award No.~DE-SC0013470. 

\bibliographystyle{elsarticle-num}
\bibliography{hqqd}

\end{document}